\documentclass[aps,pra,reprint,superscriptaddress,showpacs]{revtex4-1}

\usepackage{amsmath,amssymb,graphicx,units,ulem}
\usepackage[usenames,dvipsnames]{color}

\newcommand{\bra}[1]{\left\langle #1 \right\rvert}
\newcommand{\ket}[1]{\left\lvert #1 \right\rangle}

   \newcommand{\up}{\uparrow}
\newcommand{\dn}{\downarrow} 
 
\newcommand{\F}{\mathrm{F}} \DeclareMathOperator{\Li}{Li}

\begin{document}
\title{Coherent quench dynamics in the one-dimensional Fermi-Hubbard model}

\author{Deepak Iyer}\affiliation{Department of Physics, 
  Pennsylvania State University, University Park, PA 16802, USA}

\author{Rubem Mondaini}\affiliation{Department of Physics, 
  Pennsylvania State University, University Park, PA 16802, USA}

\author{Sebastian Will}\affiliation{Department of Physics,
  Massachusetts Institute of Technology, Cambridge, MA 02139, USA}

\author{Marcos Rigol} \affiliation{Department of Physics, 
  Pennsylvania State University, University Park, PA 16802, USA}

\begin{abstract}
  Recently, it has been shown that the momentum distribution of a
  metallic state of fermionic atoms in a lattice Fermi-Bose mixture
  exhibits coherent oscillations after a global quench that suppresses
  tunneling.  The oscillation period is determined by the Fermi-Bose
  interaction strength.  Here we show that similar coherent dynamics, but with a different functional form, occurs in
  the fermionic Hubbard model when we quench a noninteracting metallic
  state by introducing a Hubbard interaction and suppressing
  tunneling.  The period is determined primarily by the interaction
  strength. Conversely, we show that one can accurately determine the
  Hubbard interaction strength from the oscillation period, taking into
  account corrections from any small residual tunneling present in the
  final Hamiltonian. Such residual tunneling shortens the period and
  damps the oscillations, the latter being visible in the Fermi-Bose
  experiment.
\end{abstract}
\maketitle

\paragraph{Introduction.}
The Hubbard model is one of the simplest models used to describe
interacting electrons in solid state materials \cite{hubbard63}. It
describes spin-$\nicefrac12$ fermions hopping between adjacent sites
on a lattice. Opposite-spin fermions interact when they are both
present at a site. The model exhibits an interaction-driven
metal-insulator transition (Mott transition), and captures physics
of strong correlations that is believed to play a fundamental role in
high-temperature superconductivity
\cite{dagotto_review_94,imada_fujimori_98,lieb03}. The Mott transition
has already been observed at relatively high temperatures with
ultracold fermionic atoms loaded in optical lattices
\cite{jordens_strohmaier_08,schneider_hackermuller_08}.  Although
achieving lower temperatures remains an experimental challenge,
ultracold fermionic systems provide a promising venue to understand
the low-temperature phases of the Hubbard model
\cite{esslinger_review_10}.

On a different front, ultracold-atom experiments have begun the
exploration of far-from-equilibrium dynamics in isolated many-body
quantum systems.  Among many remarkable phenomena, it has been
possible to observe collapse and revival of matter waves with
Bose-Einstein condensates in optical lattices \cite{greiner02,will11},
coherent quench dynamics of a Fermi sea in a Fermi-Bose mixture
\cite{will14}, nonthermal behavior in near-integrable experimental
regimes \cite{kinoshita06,gring_kuhnert_12}, and equilibration in
Bose-Hubbard-like systems \cite{trotzky12}. These experimental
findings have motivated a large number of theoretical works seeking to
characterize and understand nonequilibrium dynamics in quantum systems
\cite{cazalilla_rigol_10,dziarmaga_10,polkovnikov11}.

We show here that the coherent quench dynamics of the fermionic
momentum distribution observed in a lattice Fermi-Bose mixture
\cite{will14} is a robust phenomenon that also occurs in purely
fermionic spin-$\nicefrac12$ systems (see Ref.~\cite{mahmud14} for
other examples of collapse and revival phenomena in bosonic and
fermionic systems).  In our study, we focus on (noninteracting)
metallic initial states at half-filling and their quench dynamics
driven by the interacting Hubbard model with suppressed site-to-site
tunneling. This is relevant to experiments where the optical lattice
is suddenly made very deep.  We show that such a quantum quench leads
to long-lived periodic oscillations of each fermionic spin species'
momentum distribution. The periodicity of the dynamics depends on the
strength of the onsite interaction between the fermions, with
quantifiable corrections due to any weak tunneling present in the
final Hamiltonian. An experimental measurement of the dynamics can
therefore be used to precisely obtain this interaction strength, even
in the presence of tunneling.

The coherent quench dynamics observed here relies on off-diagonal
(nonlocal) single particle correlations in the initial state and
serves as a signature of these. Furthermore, it generally occurs at
short times in the transient regime before thermalization takes place.
The latter is observed asymptotically after quenches in generic
isolated quantum systems \cite{rigol08} and, in particular, in
interaction quenches within the Hubbard model in dimensions higher than 
one \cite{eckstein_kollar_09}.  We present our results in the context of
the one-dimensional (1D) Hubbard model \cite{lieb68,lieb03,essler05}.
Although this model has some fundamental differences from its higher
dimensional versions (e.g., it is integrable, which means that it 
does not thermalize at long times), we do not expect these
differences to qualitatively modify our main results \footnote{At least 
for bipartite lattices.}, which are restricted to the short time dynamics.

Without loss of generality, we focus on the time evolution of
the momentum distribution of one of the fermion spin species. First, we
present analytical results for the case when the tunneling in the
final Hamiltonian is zero, where we find that the momentum
distribution oscillates in time with a period governed by the
interaction $U$ (this is not expected to change in higher
dimensions). Next, we  discuss numerical results in the case where a
finite, but small, tunneling remains after the quench. We analyze how
this modifies the period of the oscillations and leads to damping, and
discuss how one can nevertheless accurately extract the
interaction strength.  Related work in the context of the
Bose-Hubbard model was carried out in Ref.~\cite{wolf_hen_10_51}.

\paragraph{Analytical Results.}
The Hamiltonian for the Hubbard model in a periodic one-dimensional lattice is
\begin{equation}
  \label{eq:hubbard-ham}
  \hat{H} = \sum_{j=1}^{L}\left[\sum_{\sigma=\up,\dn}
    \left\{-t\left(\hat{c}^{\sigma\dag}_{j}\hat{c}^{\sigma}_{j+1} + 
        {\rm H.c.}\right)
    \right\}
    + U \hat{n}^{\up}_{j}\hat{n}^{\dn}_{j}\right],
\end{equation}
where $\hat{c}^{\sigma\dag}_{j}$($\hat{c}^{\sigma}_{j}$) creates
  (annihilates) a fermion with (pseudo-)spin $\sigma$ (denoted by
  $\up$ or $\dn$) at site $j$, $\hat{n}^{\sigma}_{j} =
  \hat{c}^{\sigma\dag}_{j}\hat{c}^{\sigma}_{j}$, $L$ is the number of
lattice sites, and
  $\hat{c}^{\sigma}_{L+1}\equiv\hat{c}^{\sigma}_{1}$ sets periodic
  boundary conditions.  We start with an initial metallic state and
quench the tunneling to zero.  We compute the momentum distribution
$n^{\sigma}_{k}(\tau) \equiv \langle
\hat{c}^{\sigma\dag}_{k}\hat{c}^{\sigma}_{k}\rangle$ as a function of
the time $\tau$ after the quench. $\hat{c}^{\sigma\dag}_{k} \equiv
\sum_{j=1}^{L}e^{\imath k a j}\hat{c}^{\sigma\dag}_{j}/\sqrt{L}$
creates a fermion with spin-$\sigma$ and momentum $k$. Using the
results for $n^{\sigma}_{k}(\tau)$, we calculate the visibility ${\cal
  V}^{\sigma}(\tau) = \int_{-k_{0}}^{k_{0}}dk\,
n^{\sigma}_{k}(\tau)$.  It measures the number of fermions with
spin $\sigma$ in the region $[-k_{0},k_{0}]$ of the Brillouin
zone. ${\cal V}^{\sigma}(\tau)$ was used in the experiments in
Ref.~\cite{will14} to characterize the time evolution of the momentum
distribution after the quench.

The initial Hamiltonian has $t_{i}=1, \, U_{i}=0$ and the final
Hamiltonian has $t_{f}=0,\,U_{f}=U$. For $N^{\up}$ and $N^{\dn}$
fermions with up and down spins respectively, the initial state (the
ground state of the initial Hamiltonian) is a Fermi sea
\begin{equation}
  \label{eq:psi-0}
  \ket{\psi_{0}} = \prod_{i=1}^{N^{\up}}\hat{c}^{\up\dag}_{k^{\up}_{i}}
  \prod_{j=1}^{N^{\dn}}\hat{c}^{\dn\dag}_{k^{\dn}_{j}}\ket{0}.
\end{equation}
In Eq.~\eqref{eq:psi-0}, $k^{\sigma}_{j}=\pm 2\pi j/(a L)$,
$j=0,1,\ldots,(N^{\sigma}-1)/2$ for odd $N^{\sigma}$, and $a$ is the
lattice spacing. For even $N^{\sigma}$, there is a degeneracy in the
ground state due to a partially filled momentum shell. In the
analytical calculations, we assume that $N^{\sigma}$ is odd (i.e., 
fully filled momentum shells) and take
the thermodynamic limit at the end. The state at time $\tau$ after the
quench is obtained via the action of the time evolution operator
$e^{-\imath\hat{H}\tau}$, where $\hat{H}$ now contains only the
interaction term (we set $\hbar=1$)
\begin{multline}
  \label{eq:psi-t}
  \ket{\psi(\tau)} =
  L^{-\frac{N^{\up}+N^{\dn}}{2}}\sum_{\{r^{\sigma}_{j}\}}\exp\left[{-\imath\tau
      U\sum_{j=1}^{N^{\up}}
      \sum_{l=1}^{N^{\dn}}\delta_{r^{\up}_{j}r^{\dn}_{l}}}\right]\\
  \times \exp\left[{\sum_{j,\sigma} \imath
      k^{\sigma}_{j}r^{\sigma}_{j}}\right]
  \prod_{i=1}^{N^{\up}}\hat{c}^{\up\dag}_{r^{\up}_{i}}
  \prod_{j=1}^{N^{\dn}} \hat{c}^{\dn\dag}_{r^{\dn}_{j}} \ket{0}.
\end{multline}
Here, $r^{\sigma}_{j}$ denotes the positions of the fermions in the
lattice, $\sum_{\{r^{\sigma}_{j}\}}$ implies a sum over all lattice sites
for each $j=1,\ldots,N^{\sigma}$, and $\sigma=\up,\dn$. $\delta_{r^{\up}_{j}r^{\dn}_{l}}$
is a Kronecker $\delta$ function.
As mentioned
earlier, the postquench dynamics is due to off-diagonal single
particle correlations present in the initial state, which evolve
in time.  Without loss of generality, we calculate this quantity
explicitly for the spin-up fermions, $\langle
\hat{c}^{\up\dag}_{m}\hat{c}^{\up}_{n}\rangle$.  The expectation value is taken in
the state at time $\tau$ [Eq.~\eqref{eq:psi-t}]. The calculation has
to be carried out separately for $m=n$ and $m\neq n$.  We first obtain
the fermion overlaps
$\bra{0}\left[\prod_{\sigma=\up,\dn}\prod_{j=1}^{N^{\sigma}}
  \hat{c}^{\sigma}_{r^{\sigma}_{j}}\right] \hat{c}^{\up\dag}_{m}\hat{c}^{\up}_{n}
\left[\prod_{\sigma=\up,\dn}\prod_{j=1}^{N^{\sigma}}
  \hat{c}^{\sigma\dag}_{r^{\sigma}_{j}}\right]\ket{0}$ as determinants
of $\delta$ functions (see Supplementary Material). After summing over
the $\delta$ functions and simplifying the time-dependent exponents, we
obtain for $m\neq n$
\begin{multline}
  \bra{\psi(\tau)}c^{\up\dag}_{m}c^{\up}_{n}\ket{\psi(\tau)} =
  L^{-N^{\dn}-1}
  \left[ \sum_{l=1}^{N^{\up}} e^{\imath k^{\up}_{l}(n-m)}\right] \\
  \times \sum_{Q}{\rm sgn}(Q)\prod_{j=1}^{N^{\dn}}
  \Bigg[\delta_{Q_{j}j}+
  (e^{-\imath\tau U}-1) e^{\imath (k^{\dn}_{j}-k^{\dn}_{Q_{j}})n} +\\
  + (e^{\imath\tau U}-1) e^{\imath (k^{\dn}_{j}-k^{\dn}_{Q_{j}})m}\Bigg],
\end{multline}
where $Q$ are permutations over $\{1,\ldots,N^{\dn}\}$.  The sum over
permutations of the product in the brackets is essentially a
determinant. It can be evaluated explicitly using the matrix
determinant lemma (see Supplementary Material). For $m=n$, $\langle
\hat{c}^{\up\dag}_{m}\hat{c}^{\up}_{m}\rangle$ is the mean site occupation, 
which is constant in time. Its value is $n^{\up}\equiv N^{\up}/L$.
We finally convert the sums over momenta to
integrals by taking the thermodynamic limit to get
\begin{multline}
  \label{eq:dens_mat}
  \bra{\psi(\tau)}c^{\up\dag}_{m}c^{\up}_{n}\ket{\psi(\tau)} =
  (1-\delta_{mn})\frac{\sin[\pi n^{\up}(m-n)]}{\pi(m-n)} \\ \times
  \Bigg[1+ 2 n^{\dn}(n^{\dn}-1)(1-\cos U\tau) - \\ - 2(1-\cos
  U\tau)\frac{\sin^{2}[\pi n^{\dn}(m-n)]}{\pi^{2}(m-n)^{2}} \Bigg] +
  n^{\up}\delta_{mn}\,.
\end{multline}
It can be verified that at $\tau=0$ we recover the single-particle
correlations of free fermions for $m\neq n$ and the site occupancies
for $m=n$. Equation \eqref{eq:dens_mat} already hints at the occurrence of 
coherent oscillations of the momentum distribution in time. 
Notice the presence of terms proportional to $\cos U\tau$. 
For comparison, in the Fermi-Bose case, one obtains
an exponential of a cosine of $U\tau$ \cite{will14}. This means that while 
on dimensional grounds the time scale for oscillations must be
proportional to $1/U$, the functional form of the time dependence
is nontrivial and depends on the system being considered.

By Fourier transforming Eq.~\eqref{eq:dens_mat}, we obtain the momentum
distribution function. The time evolution of the occupation of the
$k=0$ mode for $n^{\up}=n^{\dn}=\nicefrac{1}{2}$, i.e., at
half-filling, has the following particularly simple form:
\begin{equation}
  \label{eq:nk0}
  n^{\text{half-filling}}_{k=0}(\tau) = 1-\frac38 (1-\cos U\tau).
\end{equation}
By integrating the momentum distribution in the region $[-k_{0},k_{0}]$,
we obtain the visibility
\begin{equation}
  \label{eq:visibility}
  {\cal V}(\tau) = \frac{k_{0}}{\pi\nu} + 2g(k_{0},\nu)(1-\cos U\tau),
\end{equation}
where $g(k_{0},\nu)$ is composed of polylog functions (see
Supplementary Material for details), $\nu\equiv n^{\up,\dn}=
N^{\up,\dn}/{L}$ is the filling fraction (we assume $N^\up=N^\dn$ such that
${\cal V}^{\up}={\cal V}^{\dn}\equiv{\cal V}$), and $k_{0}\leq
\pi\nu$.  As a check, $g(k_{0},1) = 0$ as expected, because for a fully
filled Brillouin zone no dynamics is possible.

\begin{figure}[!tb]
  \centering
  \includegraphics[width=0.47\textwidth]{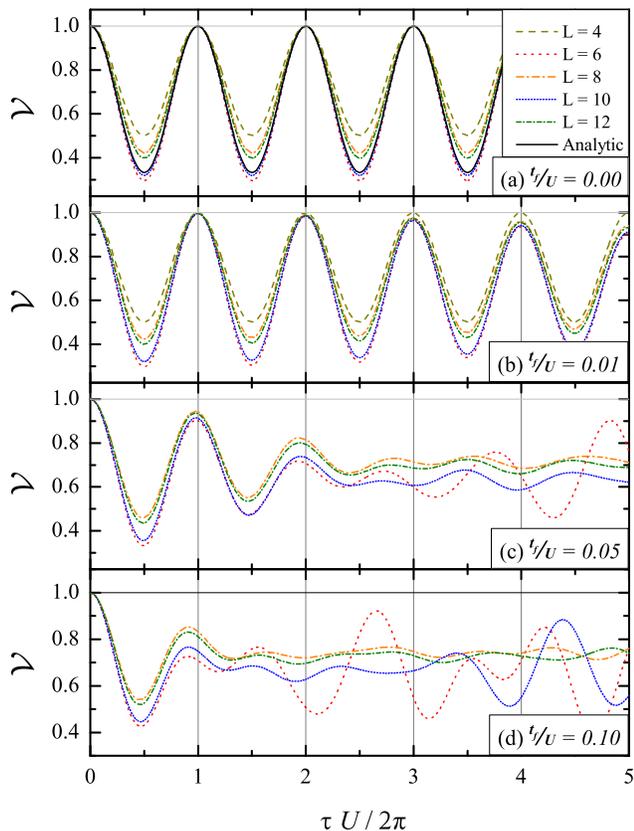}
  \caption{(Color online) Visibility as a function of time for a
    half-filled metallic initial state and different values of the
    final hopping amplitude $t_f$ [(a)--(d)].  We take $k_0$ to be the
    Fermi momentum in the initial state. The curves in each panel
    correspond to different system sizes $L$ ($4,\,6,\,8,\,10$ and
    12). Panels (a)--(d) show that there is a decrease in the revival
    time and an increase in damping, as the hopping amplitude $t_f$
    increases. The solid (black) curve in panel (a) depicts the
    analytical result for $t_f=0$ in the thermodynamic limit
    [Eq.~\eqref{eq:visibility}]. Panel (a) shows that the systems with
    $L=4$, 8, and 12 exhibit the largest finite-size effects. Also,
    note that with increasing system size they approach the
    thermodynamic limit result from above, while those with $L=6$ and
    10 approach the thermodynamic limit result from below. The case $L=4$ is not displayed in panels (c) and (d) for clarity.
    Note that panels (c) and (d) show additional revivals for some system sizes. 
    These are due to finite-size effects. All quantities plotted are dimensionless}
  \label{fig:vis-time}
\end{figure}

\paragraph{Exact Diagonalization Results.}
In what follows, we use full exact diagonalization to understand how
the analytical results in the absence of tunneling in the final
Hamiltonian are modified in the presence of a finite, but small,
tunneling amplitude. We study lattices of length
$L=4,\,6,\,8,\,10$, and 12 at half-filling. For lattice sizes $L=4m$
($m=1,2,\ldots$), the initial ground state is four-fold degenerate due to
partially filled momentum shells in the noninteracting Fermi sea. We use translation and parity
symmetries and focus on the even parity sector within the total
quasimomentum $k=0$ sector, where the ground state is not
degenerate. The resulting reduction in the size of the
relevant Hilbert space
allows us to study the exact many-body dynamics in sufficiently large systems for
arbitrarily long times.

\begin{figure}[!tb]
  \centering
  \includegraphics[width=0.47\textwidth]{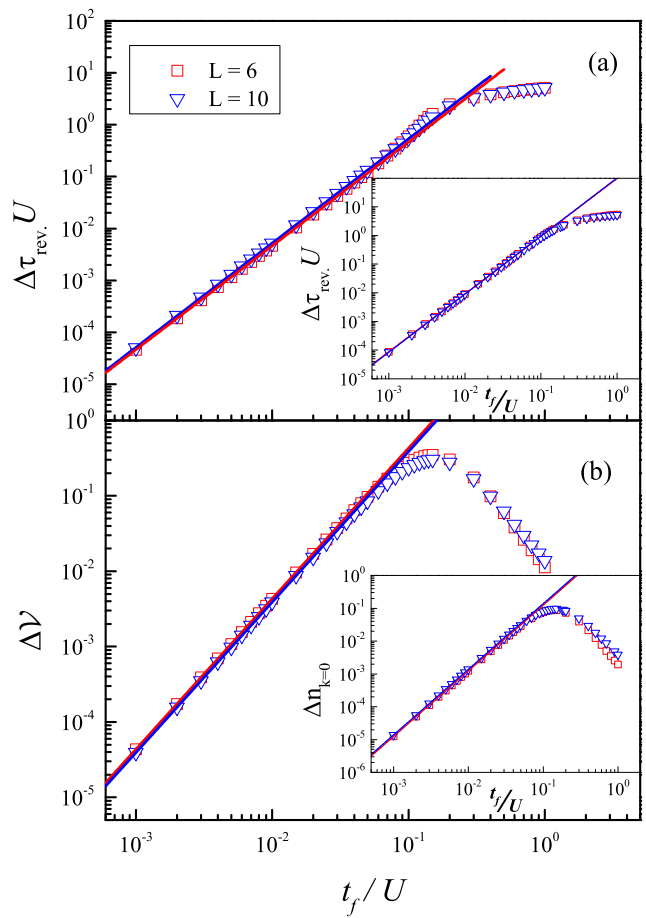}
  \caption{(Color online) (a) Absolute value of the shift in the
    visibility revival time as a function of the final hopping
    $t_f$. (b) Damping (defined as the absolute value of the change of
    ${\cal V}$ at the first revival, from the $t_f=0$ result) as a
    function of $t_f$. The insets show the corresponding results for
    the zero-momentum occupation of one of the species
    ($n_{k=0}$). All panels display results for systems with $L=6$ and
    $10$, as they exhibit the smallest finite-size effects. All quantities plotted are dimensionless.}
  \label{fig:rev_damping-t}
\end{figure}

In Fig.~\ref{fig:vis-time}, we show exact diagonalization results for
the visibility as a function of time for different values of $t_f$ 
and for the five system sizes studied. Figure~\ref{fig:vis-time}(a)
depicts results for $t_f=0$, where we also include the analytical
results in the thermodynamic limit [Eq.~\eqref{eq:visibility}]. A
comparison between the exact diagonalization results and the analytic
ones makes apparent that the systems with $L=4m$ ($L=4$, 8, and 12,
for $m=1$, 2 and 3, respectively), which correspond to partially filled 
momentum shells in the noninteracting Fermi sea, suffer from stronger 
finite-size effects than those with $L=4m+2$ ($L=6$ and 10, for $m=1$ 
and 2, respectively), which correspond to completely filled 
momentum shells in the noninteracting Fermi sea. However, with increasing 
system size, they all seem to approach the analytic prediction in the 
thermodynamic limit.

As one moves away from the ideal $t_f=0$ case and increases $t_{f}$,
two effects are clearly visible in our results for the visibility in
Fig.~\ref{fig:vis-time}: the time it takes for the system to
have the first revival decreases, and the maximum value of the
visibility at the first revival decreases, i.e., damping
increases. This is because in the presence of finite tunneling
  the local occupations are not good quantum numbers. Their change with
  time leads to decoherence in the many-body dynamics and,
  consequently, to damping of the oscillations. A finite
small tunneling can be thought of as a perturbation to the $t_{f}=0$
case. Hence, all nonconserved quantities and parameters will exhibit
  perturbative corrections proportional to powers of $t_{f}/U$.
For the largest tunneling amplitudes shown, 
$t_{f}/U=0.05$ [Fig.~\ref{fig:rev_damping-t}(c)] and 
$t_{f}/U=0.1$ [Fig.~\ref{fig:rev_damping-t}(d)], 
finite-size effects lead to large revivals of the visibility 
after a few oscillation periods. They also lead to sizable 
differences between the values of ${\cal V}$ even at the first 
revival, while the time of the first revival is barely affected 
by finite-size effects. As $t_{f}/U$ increases, deviations from 
periodic dynamics become apparent after the first oscillation
periods [see, e.g., Fig.~\ref{fig:rev_damping-t}(d)].  

In Fig.~\ref{fig:rev_damping-t}, we study the change in the revival
time (by which we mean the time of the first revival) and 
damping as a function of the final tunneling $t_{f}$, starting 
with very small values of $t_f$.  We
find that the deviation in the revival time from the value at
$t_{f}=0$ scales as $t_{f}^2$ if $U$ is unchanged
[Fig.~\ref{fig:rev_damping-t}(a)]. This is straightforward to
understand. The revival time $\tau_{\rm rev}$ has a functional form
$\tau_{\rm rev}(t_f,U) = 2\pi T(t_{f}/U)/U$, $T$ being some dimensionless
function of $t_{f}/U$. For small $t_{f}/U$, a perturbative expansion
of $T$ has a quadratic subleading term (the leading term being 1) ---
a linear term is not allowed since the Hubbard model in a bipartite
lattice is invariant under a change $t\to -t$.  By a fit to the
numerical data for $L=10$, we find that $\Delta\tau_{\rm
  rev}\equiv\tau_{\rm rev} (0,U)- \tau_{\rm rev} (t_{f},U)= 2\pi
Ct_{f}^{2}/U^{3}$, with $C=8.7 \pm 0.1$.  Similarly, we find that
$\Delta {\cal V}\equiv {\cal V}_\text{max}(0,U)-{\cal
  V}_\text{max}(t_{f},U)= Dt_{f}^{2}/U^{2}$ with $D=38.46\pm 0.04$. By
${\cal V}_\text{max}(0,U)$ and ${\cal V}_\text{max}(t_{f},U)$, we
mean the maximum of the visibility in the first revival for $t_{f}=0$
and $t_{f}\neq0$, respectively. In Fig.~\ref{fig:rev_damping-t},
finite-size effects can be seen to be slightly larger for $\Delta
{\cal V}$ than for $\Delta\tau_{\rm rev}$ so the results obtained for
the latter are expected to be closer to the thermodynamic limit
result.

If one studies the dynamics of the occupation of the $k=0$ mode of one
of the spin species ($n_{k=0}$), the results obtained are
qualitatively similar to those for the visibility [see the insets in
Fig.~\ref{fig:rev_damping-t} and Eq.~\eqref{eq:nk0}], which means that
such an observable can also be used in the experiments to study
collapse and revival phenomena in fermionic systems.

Remarkably, one can accurately determine the on-site interaction
strength $U$ in an experiment that has a small finite value of $t_{f}$
by using the measured revival time.  First, the value of
$t_{f}$ can be calculated from the known experimental lattice
parameters \cite{zwerger03}.  Since the revival time is given by
$\tau_{\rm rev} = (2\pi/U)[1-Ct_{f}^{2}/U^{2}]$, the experimentally
measured value of $\tau_{\rm rev}$, in combination with the calculated
value of $t_{f}$ and the result obtained here for $C$ (or more precise
ones which could be obtained, e.g., using time-dependent density
matrix renormalization group \cite{schollwock_review_05}) allows one
to obtain $U$ by solving the cubic equation $\tau_{\rm rev} U^{3}/2\pi
- U^{2}+C t_{f}^{2}=0$. This can also be done for bosonic systems
\cite{wolf_hen_10_51}.

\paragraph{Summary.}
We have shown that the momentum distribution function of a spin-$1/2$
metallic system exhibits coherent oscillations after a quench to a
finite interaction strength and suppressed tunneling. Similar to the
Fermi-Bose case \cite{will14}, nontrivial off-diagonal single-particle
correlations in the initial state and on-site interactions in the
final Hamiltonian are responsible for the dynamics. Experimental
observation of such dynamics would therefore provide evidence
for those off-diagonal correlations.  We have obtained analytical
results for $t_f=0$ in the thermodynamic limit, and compared them to
those obtained using full exact diagonalization of finite
systems. This allowed us to gauge finite-size effects in the exact
diagonalization calculations. The results for $L=10$ were found to be
closest to those in the thermodynamic limit.  Using exact
diagonalization, we showed that small finite residual tunneling after
the quench causes damping of the oscillations and modifies the revival
time. We argued that using the measured period of the first
oscillation from experimental data, and our (or others) theoretical
results for the constant $C$, one can obtain the interaction strength
very accurately.

\begin{acknowledgments}
  This work was supported by the U.S. Office of Naval Research
  (D.I. and M.R.), by the National Science Foundation Grant
  No.~PHY13-18303 (R.M. and M.R.) and by CNPq (R.M.).
\end{acknowledgments}

\bibliography{fermicr}

\onecolumngrid
\section*{Supplementary material}
We show details of the calculations leading to
Eq.~\eqref{eq:visibility}.  From the state at finite-time $\tau$ after
the quench, we calculate the single-particle density matrix $\langle
c^{\up\dag}_{m} c^{\up}_{n}\rangle$.  For $m\neq n$, we have
\begin{equation}
  \begin{split}
    &\bra{\psi(\tau)}c^{\up\dag}_{m}c^{\up}_{n}\ket{\psi(\tau)} =
    \sum_{\{r'^{\up,\dn}_{j}\}}\sum_{\{r^{\up,\dn}_{j}\}} e^{\sum_{j}
      \imath k^{\up}_{j}(r^{\up}_{j}-r'^{\up}_{j})+\imath
      k^{\dn}_{j}(r^{\dn}_{j}-r'^{\dn}_{j})} e^{-\imath\tau U
      \sum_{i,j}(\delta_{r^{\up}_{i}r^{\dn}_{j}} -
      \delta_{r'^{\up}_{i}r'^{\dn}_{j}})} \bra{0}
    \prod_{j=1}^{N}c^{\up}_{r'^{\up}_{j}}c^{\dn}_{r'^{\dn}_{j}}
    c^{\up\dag}_{m}c^{\up}_{n}
    \prod_{j=1}^{N}c^{\up\dag}_{r_{j}}c^{\dn\dag}_{r_{j}}\ket{0}\\
    &= \sum_{\{r'^{\up,\dn}_{j}\}}\sum_{\{r^{\up,\dn}_{j}\}}
    e^{\sum_{j} \imath k^{\up}_{j}(r^{\up}_{j}-r'^{\up}_{j})+\imath
      k^{\dn}_{j}(r^{\dn}_{j}-r'^{\dn}_{j})} e^{-\imath\tau
      U\sum_{i,j}(\delta_{r^{\up}_{i}r^{\dn}_{j}} -
      \delta_{r'^{\up}_{i}r'^{\dn}_{j}})}
    \left[\sum_{P}\sigma^{P}\prod_{j=1}^{N}
      \delta_{r^{\dn}_{j},r'^{\dn}_{P_{j}}} \right]
    \left[\sum_{Q}\sum_{l=1}^{N} \sigma^{Q}\delta_{n
        r^{\up}_l}\delta_{m r'^{\up}_{Q_{l}}}
      \prod_{i\neq l}\delta_{r'^{\up}_{Q_{i}} r^{\up}_{i}}\right]\\
    &= \sum_{\{r'^{\up}_{j},r^{\up}_{j}\}}\sum_{\{r^{\dn}_{j}\}}
    \sum_{P,Q}\sum_{l}\sigma^{P}\sigma^{Q} e^{\sum_{j} \imath
      (k^{\up}_{j}r^{\up}_{j}-k^{\up}_{Q_{j}}r'^{\up}_{Q_{j}})+\imath
      (k^{\dn}_{j}-k^{\dn}_{P_{j}})r^{\dn}_{j}} e^{-\imath\tau U
      \sum_{i,j}(\delta_{r^{\up}_{i}r^{\dn}_{j}} -
      \delta_{r'^{\up}_{i}r^{\dn}_{P_{j}}})} \left[\delta_{n
        r^{\up}_l}\delta_{m r'^{\up}_{Q_{l}}}
      \prod_{i\neq l}\delta_{r'^{\up}_{Q_{i}} r^{\up}_{i}}\right]\\
    &= \sum_{l}\sum_{\{r^{\up}_{j},j\neq
      l\}}\sum_{\{r^{\dn}_{j}\}}\sum_{P,Q}\sigma^{P}\sigma^{Q} \left[
      e^{\sum_{j\neq l} \imath (k^{\up}_{j}-k^{\up}_{Q_{j}} )r^{\up}_{j} +
        \imath (k^{\dn}_{j}-k^{\dn}_{P_{j}})r^{\dn}_{j}} e^{-\imath\tau U
        \sum_{i\neq l,j}(\delta_{r^{\up}_{i}r^{\dn}_{j}} -
        \delta_{r^{\up}_{Q_{i}}r^{\dn}_{P_{j}}})}\right]\\
    &\qquad\qquad\times e^{\imath  (k^{\up}_{l}n-k^{\up}_{Q_{l}}m) +\imath
      (k^{\dn}_{l}-k^{\dn}_{P_{l}})r^{\dn}_{l}} e^{-\imath\tau
      U\sum_{j}(\delta_{n r^{\dn}_{j}} - \delta_{m r^{\dn}_{P_{j}}})}
  \end{split}
\end{equation}
Due to the double sum in the time-dependent exponent, a dramatic
simplification occurs:
\begin{equation}
  \label{eq:5}
  \begin{split}
    &\bra{\psi(\tau)}c^{\up\dag}_{m}c^{\up}_{n}\ket{\psi(\tau)} =
    \sum_{Q}\sigma^{Q}\sum_{l}\sum_{\{r^{\up}_{j},j\neq
      l\}}\sum_{\{r^{\dn}_{j}\}}\sum_{P}\sigma^{P} \left[ \prod_{j\neq
        l} e^{\imath  (k^{\up}_{j}-k^{\up}_{P_{j}} )r^{\up}_{j} + \imath 
        (k^{\dn}_{j}-k^{\dn}_{P_{j}})r^{\dn}_{j}}
    \right]\\
    &\hspace{0.5\textwidth} \times e^{\imath 
      (k^{\up}_{l}n-k^{\up}_{P_{l}}m) +i
      (k^{\dn}_{l}-k^{\dn}_{P_{l}})r^{\dn}_{l}}
    e^{-\imath\tau U\sum_{j}(\delta_{n r^{\dn}_{j}} - \delta_{m r^{\dn}_{j}})}\\
    &= \sum_{Q}\sigma^{Q}L^{N^{\up}-1} \sum_{l=1}^{N^{\up}} e^{\imath 
      k^{\up}_{l}(n-m)}\prod_{j=1}^{N^{\dn}}\sum_{r^{\dn}_{j}} e^{\imath 
      (k^{\dn}_{j}-k^{\dn}_{Q_{j}})r^{\dn}_{j}} e^{-\imath\tau U(\delta_{n
        r^{\dn}_{j}}
      - \delta_{m r^{\dn}_{j}})}\\
    &= \sum_{Q}\sigma^{Q}L^{N^{\up}-1} \sum_{l=1}^{N^{\up}} e^{\imath 
      k^{\up}_{l}(n-m)}\prod_{j=1}^{N^{\dn}}\sum_{r^{\dn}_{j}} e^{\imath 
      (k^{\dn}_{j}-k^{\dn}_{Q_{j}})r^{\dn}_{j}}[1+ (e^{-\imath\tau
      U}-1)\delta_{n r^{\dn}_{j}}
    + (e^{\imath\tau U}-1)\delta_{m r^{\dn}_{j}}]\\
    &= L^{N^{\up}-1}\left[ \sum_{l=1}^{N^{\up}} e^{\imath 
        k^{\up}_{l}(n-m)}\right]
    \sum_{Q}\sigma^{Q}\prod_{j=1}^{N^{\dn}}
    \left[\frac{\sin[(k^{\dn}_{j}-k^{\dn}_{Q_{j}})L/2]}{\sin[(k^{\dn}_{j}-k^{\dn}_{Q_{j}})/2]}+
      (e^{-\imath\tau U}-1) e^{\imath  (k^{\dn}_{j}-k^{\dn}_{Q_{j}})n} +
      (e^{\imath\tau U}-1)
      e^{\imath  (k^{\dn}_{j}-k^{\dn}_{Q_{j}})m}\right]\\
    &= L^{N^{\up}-1}\left[ \sum_{l=1}^{N^{\up}} e^{\imath 
        k^{\up}_{l}(n-m)}\right]L^{N^{\dn}} \left[1+
      2\left\{\frac{N^{\dn}(N^{\dn}-1)}{L^{2}} -\frac{N^{\dn}}{L}
      \right\}(1-\cos U\tau) -2(1-\cos U\tau)L^{-2}
      \sum_{\imath ,j=1}^{N^{\dn}}e^{\imath(m-n)(k^{\dn}_{i}-k^{\dn}_{j})} \right]\\
    &= L^{N^{\up}+N^{\dn}-1} \left[ \sum_{l=1}^{N^{\up}} e^{\imath 
        k^{\up}_{l}(n-m)}\right] \left[1+ 2 n^{\dn}(n^{\dn}-1)(1-\cos
      U\tau) -2(1-\cos
      U\tau)L^{-2}\sum_{i,j=1}^{N^{\dn}}e^{\imath(m-n)(k^{\dn}_{i}-k^{\dn}_{j})}\right]
  \end{split}
\end{equation}
In the second to last line, we have evaluated the sum over
permutations $Q$ -- it is essentially the determinant of a matrix
$\mathbf{A}$ whose elements are
\begin{equation}
  \label{eq:12}
  A_{ij} = L\delta_{ij}+ \alpha e^{\imath (k^{\dn}_{i}-k^{\dn}_{j})n} + 
  \alpha^{*}e^{\imath  (k^{\dn}_{i}-k^{\dn}_{j})m}
\end{equation}
where we have taken into account that in the limit of large $L$ the
first term vanishes unless $k^{\dn}_{i}=k^{\dn}_{j}$, and set $\alpha
= e^{-\imath U\tau}-1$.  Introducing the following vectors
\begin{equation}
  \label{eq:13}
  \begin{split}
    [\mathbf{u}]_{j} &= \alpha e^{\imath k^{\dn}_{j}n}, \qquad
    [\mathbf{v}]_{j} = e^{-\imath k^{\dn}_{j}n},\\
    [\mathbf{u}]'_{j} &= \alpha^{*} e^{\imath k^{\dn}_{j}m}, \qquad
    [\mathbf{v}]'_{j} = e^{-\imath k^{\dn}_{j}m},
  \end{split}
\end{equation}
the matrix $\mathbf{A}$ can be written as
\begin{equation}
  \label{eq:14}
  \begin{split}
    \mathbf{A} &= L\mathbf{1} + \mathbf{u}\mathbf{v}^{T} +
    \mathbf{u}'\mathbf{v}'^{T} \\
    & = L\mathbf{1} + \begin{pmatrix}\mathbf{u} &
      \mathbf{u}'\end{pmatrix} \begin{pmatrix} \mathbf{v}^{T} \\
      \mathbf{v}'^{T}\end{pmatrix}
  \end{split}
\end{equation}
We then use the Matrix Determinant lemma (or Sylvester's theorem)
\cite{akritas96} to write the determinant of $\mathbf{A}$ as
\begin{equation}
  \label{eq:15}
  \begin{split}
    \det \mathbf{A} &= L^{N^{\dn}}\det\left[\mathbf{1} + L^{-1}
      \begin{pmatrix} \mathbf{v}^{T} \\
        \mathbf{v}'^{T}\end{pmatrix}
      \begin{pmatrix}\mathbf{u} &
        \mathbf{u}'\end{pmatrix} \right] \\
    &= L^{N^{\dn}}
    \left[(1+L^{-1}\mathbf{v}^{T}\mathbf{u})(1+L^{-1}\mathbf{v}'^{T}\mathbf{u}')
      - L^{-2}\mathbf{v}^{T}\mathbf{u}'\mathbf{v}'^{T}\mathbf{u}\right]\\
    &= L^{N^{\dn}}\left[ (1+\alpha n^{\dn})(1+\alpha^{*}n^{\dn}) -
      L^{-2}\alpha\alpha^{*}\sum_{i,j}
      e^{\imath(k^{\dn}_{i}-k^{\dn}_{j})(m-n)}
    \right] \\
    &= L^{N^{\dn}}\left[ 1+2(1-\cos U\tau)n^{\dn}(n^{\dn}-1)
      -2L^{-2}(1-\cos U\tau)\sum_{i,j}
      e^{\imath(k^{\dn}_{i}-k^{\dn}_{j})(m-n)} \right].
  \end{split}
\end{equation}
For $m=n$, the above calculation can be repeated, and results in
\begin{equation}
  \label{eq:6}
  \begin{split}
    \bra{\psi(\tau)}c^{\up\dag}_{m}c^{\up}_{n}\ket{\psi(\tau)} &=
    L^{N^{\up}-1}
    \sum_{l=1}^{N^{\up}}\prod_{j=1}^{N^{\dn}}\sum_{r^{\dn}_{j}}
    \delta_{mn}
    e^{\imath(k^{\dn}_{j}-k^{\dn}_{Q_{j}})r^{\dn}_{j}}     \\
    &= N^{\up} L^{N^{\up}+N^{\dn}-1}\delta_{mn}\,.
  \end{split}
\end{equation}
Normalizing both expressions, we finally get
\begin{multline}
  \label{eq:7}
  \bra{\psi(\tau)}c^{\up\dag}_{m}c^{\up}_{n}\ket{\psi(\tau)} =\\
  (1-\delta_{mn})L^{-1}\left[ \sum_{l=1}^{N^{\up}} e^{\imath 
      k^{\up}_{l}(n-m)}\right] \left[1+ 2 n^{\dn}(n^{\dn}-1)(1-\cos
    U\tau) -2L^{-2}(1-\cos
    U\tau)\sum_{i,j=1}^{N^{\dn}}e^{\imath(m-n)(k^{\dn}_{i}-k^{\dn}_{j})}\right]
  + n^{\up}\delta_{mn}\,.
\end{multline}
The momentum distribution is given by
\begin{equation}
  \label{eq:8}
  n^{\up}_{k} = L^{-1}\sum_{m,n}e^{\imath k(m-n)}\langle
  c^{\up\dag}_{m}c^{\up}_{n}\rangle.
\end{equation}
The sums over the momenta can be converted to integrals in the
thermodynamic limit. However, it is important that we do not change
the sum over $m,n$ to an integral (the lattice is fundamental)
\begin{equation}
  \label{eq:4}
  \begin{split}
    n_{k}(\tau) &= \sum_{m,n}e^{\imath k(m-n)} \Bigg\{ L^{-1}\left[
      \int_{-k_{\F}}^{k_{\F}}\frac{{\rm d}k^{\up}}{2\pi} e^{\imath 
        k^{\up}(n-m)}\right]
    \Bigg[1+ 2 n^{\dn}(n^{\dn}-1)(1-\cos U\tau) -2(1-\cos U\tau)\\
    &\hspace{0.6\textwidth} \times\int_{-k_{\F }}^{k_{\F}} \frac{{\rm
        d}k^{\dn}_{1}}{2\pi}\frac{{\rm d}k^{\dn}_{2}}{2\pi}
    e^{\imath(m-n)(k^{\dn}_{1}-k^{\dn}_{2})}\Bigg]\\
    &\qquad\qquad+ L^{-1}\delta_{mn}\left(n^{\up} - n^{\up} \left[1+ 2
        n^{\dn}(n^{\dn}-1)(1-\cos U\tau) -2(1-\cos
        U\tau)(n^{\dn})^{2}\right]\right) \Bigg\}\\
    &= \sum_{m,n}e^{\imath k(m-n)} \Bigg\{ L^{-1} \frac{\sin(\pi
      n^{\up}(m-n))}{\pi(m-n)} \left[1+ 2 n^{\dn}(n^{\dn}-1)(1-\cos
      U\tau) -2(1-\cos U\tau)\frac{\sin^{2}
        (\pi n^{\dn}(m-n))}{\pi^{2}(m-n)^{2}}\right]\\
    &\qquad\qquad+ L^{-1}\delta_{mn} 2n^{\dn}n^{\up}(1-\cos U\tau)\Bigg\}\\
    &= \sum_{m,n}e^{\imath k(m-n)} \Bigg\{ L^{-1} \frac{\sin(\pi
      n^{\up}(m-n))}{\pi(m-n)} \left[1+ 2 n^{\dn}(n^{\dn}-1)(1-\cos
      U\tau) -2(1-\cos U\tau)\frac{\sin^{2}
        (\pi n^{\dn}(m-n))}{\pi^{2}(m-n)^{2}}\right]\\
    &\qquad\qquad+ L^{-1}\delta_{mn} 2n^{\dn}n^{\up}(1-\cos
    U\tau)\Bigg\}.
  \end{split}
\end{equation}
The occupation of the $k=0$ momentum mode is given by
\begin{equation}
  \label{eq:2}
  \begin{split}
    n_{k=0}(\tau) &= \sum_{m,n}\Bigg\{ L^{-1} \frac{\sin(\pi
      n^{\up}(m-n))}{\pi(m-n)} \left[1+ 2 n^{\dn}(n^{\dn}-1)(1-\cos
      U\tau) -2(1-\cos U\tau)\frac{\sin^{2}
        (\pi n^{\dn}(m-n))}{\pi^{2}(m-n)^{2}}\right]\\
    &\qquad\qquad+ L^{-1}\delta_{mn} 2n^{\dn}n^{\up}(1-\cos
    U\tau)\Bigg\} \\
    &= \sum_{m^{-}}\Bigg\{ \frac{\sin(\pi n^{\up}m^{-})}{\pi m^{-}}
    \left[1+ 2 n^{\dn}(n^{\dn}-1)(1-\cos U\tau) -2(1-\cos
      U\tau)\frac{\sin^{2}(\pi
        n^{\dn}m^{-})}{\pi^{2}(m^{-})^{2}}\right]\Bigg\}
    + 2n^{\dn}n^{\up}(1-\cos U\tau)\\
    &= \left[1+ 2 n^{\dn}(n^{\dn}-1)(1-\cos U\tau) -2(1-\cos
      U\tau)f(n^{\up},n^{\dn})\right]
    + 2n^{\dn}n^{\up}(1-\cos U\tau)\\
    &= 1 + 2(1-\cos U\tau)\left[n^{\dn}(n^{\up}+n^{\dn}-1)-
      f(n^{\up},n^{\dn})\right],
  \end{split}
\end{equation}
where
\begin{equation}
  \label{eq:3}
  f(n^{\up},n^{\dn}) = (n^{\dn})^{2}n^{\up} + \frac{\imath }{4\pi^{3}}\left[2\Li_{3}(e^{-\imath\pi n^{\up}})+ 
    \Li_{3}(e^{\imath\pi(2n^{\dn}+ n^{\up})}) + \Li_{3}(e^{-\imath\pi(2n^{\dn}- n^{\up})}) - \text{c.c.}\right],
\end{equation}
and $\Li_{s}(z)\equiv \sum_{k=1}^{\infty}\frac{z^{k}}{k^{s}}$ is the
polylogarithm function. For $n^{\up}=n^{\dn}=\nicefrac12$, we get
\begin{equation}
  \label{eq:10}
  n^{\text{half-filling}}_{k=0}(\tau) = 1-\frac38 (1-\cos U\tau)
\end{equation}

To compute the visibility, for simplicity, we assume that $\nu\equiv
n^{\up}=n^{\dn}=k_{\rm F}/\pi$ and that $k_{0}<k_{\F}$. We get that,
for $k_{0} \leq \pi\nu$,
\begin{equation}
  \label{eq:9}
  \begin{split}
    {\cal V}(\tau) &= \frac{1}{n^{\up}}\sum_{m,n}
    \frac{\sin(k_{0}(m-n))}{\pi (m-n)} \Bigg\{ L^{-1} \frac{\sin(\pi
      n^{\up}(m-n))}{\pi(m-n)} \left[1+ 2 n^{\dn}(n^{\dn}-1)(1-\cos
      U\tau) -2(1-\cos U\tau)
      \frac{\sin^{2}(\pi n^{\dn}(m-n))}{\pi^{2}(m-n)^{2}}\right]\\
    &\qquad\qquad+ L^{-1}\delta_{mn} 2n^{\dn}n^{\up}(1-\cos U\tau)\Bigg\} \\
    &= \frac{k_{0}}{\pi\nu}[1+ 2 \nu(\nu-1)(1-\cos U\tau)] +2(1-\cos
    U\tau)g(k_{0},\nu)
    + 2\frac{k_{0}}{\pi}\nu^{2}(1-\cos U\tau) \\
    &= \frac{k_{0}}{\pi\nu} + 2g(k_{0},\nu)(1-\cos U\tau),
  \end{split}
\end{equation}
where $g(k_{0},\nu)$ is given by:
\begin{multline}
  \label{eq:11}
  g(k_{0},\nu) = \frac{k_{0}\nu}{\pi} + \frac{k_{0}(\nu-1)}{\pi} -
  \frac{k_{0}\nu^{2}}{\pi} +\frac{1}{8\pi^{4}\nu} \bigg[
  \Li_{4}(e^{-\imath(k_{0}-3\pi\nu)}) + \Li_{4}(e^{\imath(k_{0}-3\pi\nu)}) -
  \Li_{4}(e^{-\imath(k_{0}+3\pi\nu)}) - \Li_{4}(e^{\imath(k_{0}+3\pi\nu)}) \\
  + 3\Li_{4}(e^{-\imath(k_{0}+3\pi\nu)}) + 3\Li_{4}(e^{\imath(k_{0}+3\pi\nu)}) -
  3\Li_{4}(e^{-\imath(k_{0}-3\pi\nu)}) -3\Li_{4}(e^{\imath(k_{0}-3\pi\nu)})
  \bigg].
\end{multline}

\end{document}